\documentclass[hidelinks,12pt]{article}
\usepackage{setspace,amsmath,amssymb,amsthm,mathrsfs,sectsty,hyperref,xcolor,graphicx,geometry}
\usepackage[square,sort,comma,numbers]{natbib}
\hypersetup{
    colorlinks,
    linkcolor={red!50!black},
    citecolor={red!50!black},
    urlcolor={blue!50!green}
}
\allsectionsfont{\centering}
\begin{document}

\title{Is the statement of Murphy's Law valid?}
\author{Atanu Chatterjee\footnote{\href{mailto:atanu@smail.iitm.ac.in}{atanu@smail.iitm.ac.in}}\\
Indian Institute of Technology Madras\\
Chennai-600036, India}

\date{}

\maketitle

\begin{abstract}

\noindent Murphy's Law is $not$ a law in the formal sense yet popular science often compares it with the Second Law of Thermodynamics as both the statements point toward a more disorganized state with time. In this paper, we first construct a mathematically equivalent statement for Murphy's Law and then disprove it using the intuitive idea that energy differences will level off along the paths of steepest descent, or along trajectories of least action. 
\end{abstract}



\tableofcontents

\newpage
\section{Introduction}
\noindent Murphy's Law is not a law in the formal sense; instead it is an interesting satirical statement: \emph{If something can go wrong then it will definitely go wrong} \cite{bloch2003murphy}. The origin of the statement is unknown (even Murphy himself!), yet it is not very hard to find it relevant in various situations, either in its rudimentary form or in the form of some derivative. Popular science often relates this statement to the Second Law of Thermodynamics, which states that the entropy of the universe will always increase with time \cite{handscombe2004entropy, cengel2007introduction}. The analogy between these two seemingly very dissimilar statements seems to crop up from the idea that left to themselves, things tend to get disorganized with time and Murphy comes into the picture, albeit in a modified form, stating that bad will eventually turn worse. In the realm of philosophy these statements may seem connected, however from a scientific viewpoint they are clearly unrelated, as (\emph{i}) one is a universal law, whereas the other, a rather ambiguous statement; (\emph{ii}) the second-law strictly talks about energy dissipation in physical processes and their irreversibility, whereas Murphy's statement (`law' is a misnomer in this case) is just a popular imagination, and does not even remotely relate to any of the physical or mathematical concepts to rely upon. Surprisingly, given these discrepancies, we still find it very interesting when the law is formulated as a mathematically equivalent statement. In this paper, we try to \emph{actually} relate Murphy's statement with the Second Law of Thermodynamics, and draw inspirations from energy and entropy. We check the correctness of the statement from a scientific standpoint, by first constructing a mathematically equivalent statement using concepts from mathematical logic and probability theory. Next, we understand the physical essence of the terms `rightness' and `wrongness' from the context of energy and time, using the Second Law and Principle of Least Action. Finally, we relate our findings to check the truthfulness of Murphy's statement. We observe that what initially seemed to be completely unrelated statements - the Second Law and Murphy's statement - \emph{indeed} have some interesting tales to tell.
\section{Methodology}
\subsection{Murphy and Logic}
Murphy's statement is more of a psychological statement having philosophical connotations of `rightness' and `wrongness'. A statement like this is not only hard to evaluate but even more challenging to prove (or disprove). The first step in any proof strategy lies in the rigorous formulation of the statement which is under scrutiny, and hence it is necessary that we try to formulate this rather anecdotal philosophy into a logical one. Doing so lets us set limits to the premise of the statement, as well as to its domain of application. However, in order to proceed we need to make some underlying assumptions. The idea of using propositional logic to build a mathematically coherent statement, even before attempting to prove or disprove it, is a necessity. Whereas, assigning an additional element of probability to the statement(s) is actually non-intuitive, because (\emph{i}) Murphy's speaks about occurrence of events and in mathematics, occurrence is always associated with \emph{chance}; (\emph{ii}) the presence of the intensifier, \emph{`definitely'}, in the linguistic construction of the statement itself, further strengthens our intuition of associating the statement(s) with finite values of probability. Logically speaking, Murphy's statement can be broadly divided into two distinct components: \textbf{1}. (\emph{If}) something can go wrong, and \textbf{2}. (\emph{then}) it will definitely go wrong. In propositional logic, this can be viewed as a causal chain of statements, $p \rightarrow q$, where $p$ is the first component, $q$ the second, and $p$ causes $q$. Such an argument is known as \emph{modus ponnens} \cite{tarski1993introduction} which asserts that the occurrence of $q$ is dependent upon the occurrence of $p$, or if $p$ happens, $q$ definitely happens. It can be represented mathematically as $\frac{p \rightarrow q, \enspace p}{\therefore, \enspace q}$. It was noted earlier that the above statements $p \rightarrow q$ have an inherent probability associated with them. The presence of an intensifier, `definitely' further affirms the occurrence of the latter half of the statement, irrespective of the magnitude of the chances of occurring of the former half. Mathematically, if the probability of occurrence of $p$ is \emph{finite} (remember, the probability that $\neg p$ occurs is also finite) then $q$ occurs \emph{almost surely}, \emph{i.e.}, with probability one (note that, even though we are interested in the happening of something which is not desired). What we earlier stated as a \emph{modus ponnens}, interestingly transforms into \emph{modus tollens} once we associate finite probabilities to the occurrence of the events (\emph{modus ponnens} is the law of affirming by affirming, therefore we affirm $p$ in order to justify $q$, and Murphy's statement affirms $q$ by associating finite chances of occurrence of $p$) \cite{tarski1993introduction}. According to the Law of Contrapositive: (\emph{modus tollens}), $\frac{p \rightarrow q, \enspace \neg q}{\therefore, \enspace \neg p}$, or if probability of occurrence of $p$ is \emph{finite} then $q$ occurs \emph{almost surely} and $q$ did not occur \emph{almost surely} enforces the fact that $p$ did not have a finite chance of occurring in the first place, or $\neg q \Rightarrow \neg p$ \cite{billingsley2008probability}. This is the form of Murphy's statement that we will adhere to throughout this paper. Also, in the following sections of this paper, we will analyze Murphy's statement using fundamental laws of physics, the Least Action Principle and the Second Law of Thermodynamics. It may seem puzzling at first as to how energy and entropy come into the picture, but a deeper inspection will reveal the intimate connection that Murphy's statement holds with respect to a system, process and entropy. 
\subsection{Murphy, Thermodynamics and Action}
\noindent The connection between Murphy's statement and thermodynamics is the central theme of this paper. As proposed earlier, when we talk about Murphy's statement we refer to the probability of occurrence of any event in space and time. We thus make a transition from a system-theoretic perspective to a process-theoretic one, and once we do that we inevitably include the Second Law into the picture. Where, thermodynamics is a branch of science dealing with energy interactions with systems, energy conversion and energy degradation; it has also given birth to one of the most fundamental laws of nature, known as the Second Law or the Entropy Principle \cite{cengel2007introduction, clausius1854veranderte, planck2013treatise, lieb1999physics}. The Second Law deals with processes in nature, and acts as a standard test to justify the occurrence or nonoccurrence of any phenomenon. Every process (phenomenon) in nature can be visualized as a flow of energy. When this flow is spontaneous, \emph{i.e.}, the driving force behind the flow is the presence of gradients (like, energy, temperature, pressure, concentration, etc.), then such a flow is realized with an equivalent increase in entropy \cite{sharma2007natural, kaila2008natural, gronholm2007natural, pernu2012natural}. The flow of energy during any arbitrary process when coupled with the dimension of time gives rise to the notion of \emph{action}, formally represented as $\mathcal{A} = \int_{A \to B} (T-V)dt = \int_{A \to B} (\delta E)dt = \int_{A \to B} \mathcal{L}(p_{i},q_{i})dt$, where the pair, $(q_{i},p_i)$ is the generalized position-momentum pair, $t$ is the time, $\delta E$ is the change in energy along a specific path (in this case $A \to B$), $T$ and $V$ are the kinetic and potential energies respectively and $\mathcal{L} (p_i , q_i)$, the Lagrangian \cite{goldstein1957classical, de1751essais, euler1952methodus, lagrange1853mecanique}. There are several interpretations of the Action Principle, for example, the laws of physics when formulated in terms of the Action Principle identify any natural process as the one in which energy differences are leveled off in the least possible time (Maupertius' formulation) \cite{kaila2008natural, pernu2012natural}; among all existing possible paths, the path of least action is the one along which a natural process must proceed (Hamiltonian formulation) \cite{georgiev2002least, georgiev2012quantitative, georgiev2014mechanism, georgiev2014spontaneous, chatterjee2012action, chatterjee2013principle, chatterjee2011certain, atanu}. On the energy landscape, there is a unique trajectory that directs (energy) flows along the lines of steepest descent. A classical example of this principle can be observed in case of the Brachistochrone problem. The Principle of Least Action, thus dictates the existence of a unique path out of numerous possible trajectories (see figure 1). While some are thermodynamically feasible, others are not. From a thermodynamic perspective decrease in action is equivalent to an increase in entropy (we restrict ourselves to non self-organizing systems and processes), or along a trajectory, both the quantities can be represented along mutually orthogonal directions but with opposite signs, $\mathcal{L}=-T\mathcal{S}$ \cite{atanu, lucia2008probability, lucia2012maximum}. Murphy's statement vaguely mentions processes, but stresses more on their qualitative aspect of `rightness' and `wrongness'. It does not talk about the system, per se, undergoing the process, nor does it go into the details of the mechanism underlying the process; it basically concentrates only on the qualitative aspects of the outcomes. In order to test the correctness of the statement, it is absolutely necessary to look into the details of the process, as focusing on the finer details of the process will enable us to justify the outcomes.
\subsection{Defining a Probability Measure}
\noindent A deeper inspection is needed when we talk about processes and trajectories (pathways to achieve a particular outcome during a process). When we talk about Murphy's statement, we must invariably refer to processes and the trajectories (pathways) along which these processes occur (flow). In order to model a physical phenomenon into a thermodynamic one, we must explicitly distinguish between a system, surroundings and the set of processes that the system undergoes. In the algebraic formulation of variational thermodynamics, these terms are defined in a detailed, mathematically rigorous yet an abstract fashion \cite{atanu, lucia2008probability, lucia2012maximum, lucia1995mathematical, caratheodory1909untersuchungen, truesdell1984rational}. Roughly speaking, a system is a set of all those elements which are of current interest, and there exists a boundary that separates the system from its surroundings, where surroundings (relative to the system) refer to everything else in the universe except the system and the system boundary. A process is a change of state of a subset of system elements or the system itself in time. Mathematically, a process can be represented as a map that captures the changes incurred by the system or a subset of a system. We define a system undergoing a process by a tuple, ($\Omega , \Pi$), where the elements of $\Omega$ are the system elements/constituents/agents, and the elements of $\Pi$ are processes \cite{atanu}. We identify an event as a \emph{spontaneous} process in space-time if it proceeds along the path that reduces action and increases entropy \cite{atanu, lucia2008probability, lucia2012maximum}. The Principle of Least Action is that tool which helps us in finding that particular trajectory out of a set of uncountable possibilities along which an event must proceed. Using the standard notation, we define action for an arbitrary element in a system undergoing an arbitrary process, as $\mathcal{A}_{ij}=\int_{j}p_i dq_i$, where $i \in \mathcal{I}$, the set of all system elements in $\Omega$, and $j \in \mathcal{J}$, the set of all arbitrary processes, $\pi_j\in \Pi$ \cite{atanu}. For any arbitrary process, $\pi_j$ there exist numerous possibilities or trajectories, $\gamma_k$. We define the set $\Gamma^j$ to be the set of all such trajectories, $\gamma_k$ for the corresponding process, $\pi_j$ $\in \Pi$ and $\forall k \in \mathcal{K}$ ($\mathcal{I}$, $\mathcal{J}$, and $\mathcal{K}$ are index sets). It is now crucial to define what exactly is meant by `right' and `wrong' events as this is the central idea of Murphy's statement. We define two subsets of trajectories, $R$ and $W$, such that both $(R, W) \in \Gamma^j$. The trajectories which are thermodynamically feasible form the set, $R$, and the ones which are not, form the set, $W$. Therefore, the elements of $\Gamma^j$ are either labeled by $\gamma^{w}_k$ or $\gamma^{r}_k$ (for representation purpose only). The right trajectories are denoted by $\gamma^r_{k} \in R \subset \Gamma^j$, and the wrong ones by $\gamma^w_{k} \in \Gamma^j \setminus R \subset \Gamma^j = W$. Clearly, the subsets, $R$ and $W$ partition the set, $\Gamma^j$, so $R \cup W = \Gamma^j$ and $R \cap W = \phi$ (see figure 2). It is not very hard to see that the tuple, ($\Gamma^j , \mathcal{G}$) forms a $\sigma$-algebra\footnote{Let $\mathcal{G}$ be a collection of sets of $\Gamma^j$, then ($W$, $R$) $\in \mathcal{G}$, ($W^c$, $R^c$) $\in \mathcal{G}$, $\phi \in \mathcal{G}$, $\Gamma^j \in \mathcal{G}$, and finally $R\cup W \in \mathcal{G}$ and $R \cap W \in \mathcal{G}$. So, ($\Gamma^j , \mathcal{G}$) is a $\sigma$-algebra.} where $\mathcal{G}$ is a collection of sets of $\Gamma^j$ \cite{billingsley2008probability}. We can define a probability measure, $\mathcal{P} (\cdot)$, such that $\mathcal{P}(\Gamma^j)=1$ and $\mathcal{P}(\phi)=0$, then the triple, ($\Gamma^j ,\mathcal{G}, \mathcal{P}$) forms a probability space.
\begin{figure}[t]
\begin{center}
\includegraphics[width=0.4\textwidth]{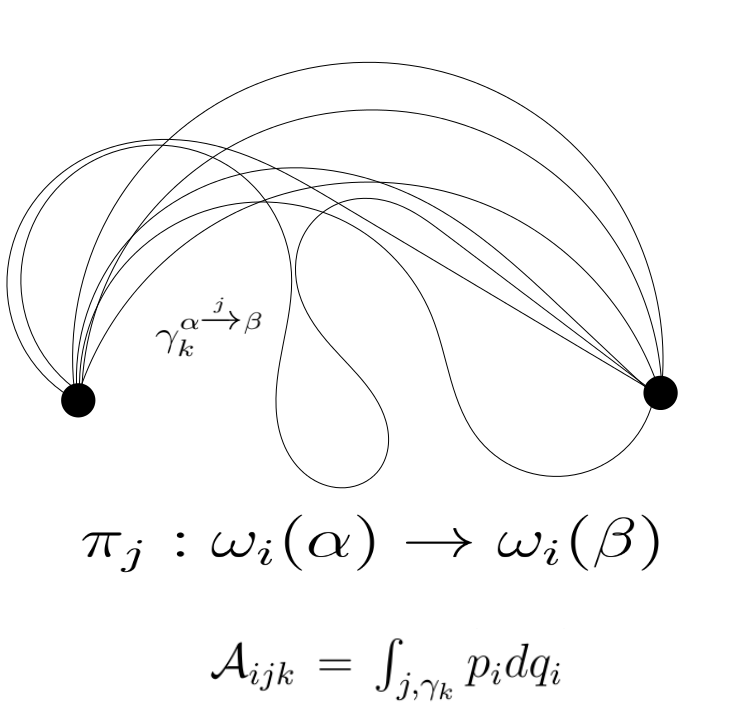}
\end{center}
\caption{Figure shows numerous trajectories of a particle $\omega_i\in\Omega$, and the action $\mathcal{A}_{ijk}$ (for $i^{th}$ particle) undergoing $j^{th}$ process along $k^{th}$ trajectory, $\gamma_k$ \cite{atanu}.}
\end{figure}
\section{Mathematical Proof}
As stated above, every element in $\Gamma^j$ is a possible trajectory, and it is not surprising that the set is unbounded because there exist infinitely many possible trajectories or sequence of trajectories for a given outcome (Feynman's path integral formulation argument) \cite{feynman2013feynman}. Furthermore, we associate some \emph{finite} (however small or large) probability to every trajectory (trajectories are basically the worldlines of a particular outcome). Thus, without any loss of generality, we can assume that the set, $W$, is \emph{countably} infinite, whereas the set, $R$, is finite. It is even more interesting to note that the set, $W$ is, in fact, \emph{uncountable}. Since the elements of the set $W$ represent all those trajectories which are thermodynamically infeasible, we can make any arbitrary combination of $\gamma^w_{k}$ (or a mix of $\gamma^w_{k}$ and $\gamma^r_{k}$, because it doesn't matter if a process is thermodynamically feasible in a later stage if it is infeasible in the very first instant!), and always come up with a new infeasible trajectory, say $\gamma^{w}_{k1}, \gamma^w_{k2}, ...$. As the set $W$ is infinite, we come up with an argument similar to Cantor's Diagonalization argument for uncountability of Reals \cite{halmos1960naive}, and prove that $W$ is uncountable because for every increment in $k$, we can always find a new element, $\gamma^w_{kt} \in W$ based on the combination of the diagonal elements (in this case, $\gamma^w_{kt}=\gamma^w_{1}\gamma^w_{3}\gamma^w_{5} ... \dotso \gamma^w_{\theta}$) and this will happen infinitely many times, \emph{i.e.},
\begin{eqnarray*}
\gamma^w_{k1}&=&\underline{\gamma^w_{1}}\gamma^w_{2}\gamma^w_{3}...\\
\gamma^w_{k2}&=&\gamma^w_{2}\underline{\gamma^w_{3}}\gamma^w_{4}...\\
\gamma^w_{k3}&=&\gamma^w_{3}\gamma^w_{2}\underline{\gamma^w_{5}}...\\
\vdots\\
\gamma^w_{kt}&=&\gamma^w_{9}\gamma^w_{7}\gamma^w_{6} ... \dotso \underline{\gamma^w_{\theta}}
\end{eqnarray*} 
For the physical explanation of the above argument, we consider the case of a chemical reaction where substrates are converted into products. In order that the reaction is thermodynamically feasible, a minimum energy barrier (energy hill) called Activation Energy, $E_a$, has to be surpassed \cite{atkins2005elements}. Those trajectories along which this energy threshold is successfully realized can be visualized as the `right' trajectories and vice-versa. Even the so called `right' trajectories have further thermodynamic constraints, so basically, we can refine our intuition and come up with a finite number of possibilities, or maybe in some exceptional cases, \emph{atmost} countable `right' possibilities and \emph{always} uncountable `wrong' possibilities. 
\subsection{Transformed Probability Space and Lebesgue Measure}
Let $\mathcal{T}_\mathcal{A}(\cdot)$ be an action map such that, $\mathcal{T}_\mathcal{A}=\frac{\mathcal{A}_{ijk}}{\max_k\mathcal{A}_{ijk}}$, in this case, $\mathcal{T}_\mathcal{A}(\gamma_k)=\frac{\mathcal{A}_{k}}{\max_k\mathcal{A}_{k}}=\frac{\int_{j,\gamma_k}p_i dq_i}{\max_k\int_{j,\gamma_k}p_i dq_i}$, and $\mathcal{T}_\mathcal{A}: \Gamma^j \rightarrow [0,1]$. The function, $\mathcal{T}_\mathcal{A}(\cdot)$, takes every trajectory, $\gamma_k$ from the set $\Gamma^j$, and calculates the action along that trajectory to the one along which the magnitude of action is maximum, which forms the transformed set, $\Gamma^{j\star}$. In fact, the map, $\mathcal{T}_\mathcal{A} (\cdot)$ is a homeomorphism\footnote{The inverse image of $\Gamma^{j\star}$ are the trajectories in $\Gamma^j$. The action map is similar to the pay-off function in Game Theory \cite{anttila2011natural} and Sigma-Profiles in multi-agent complex system modeling \cite{gershenson2011sigma}.}, $\mathcal{T}_\mathcal{A} (\cdot):\Gamma^j\rightarrow\Gamma^{j\star}$ (see figure 2). From our previous arguments, there are certain trajectories which are absolutely infeasible (although having a finite probability, however small), $\mathcal{A}_k \sim \max_k\mathcal{A}_k$ and few feasible ones, $\mathcal{A}_k \ll \max_k\mathcal{A}_k$. Thus, the transformed sample space, $\Gamma^{j\star}$ would take up all the values in the compact set, $\Gamma^{j\star}=[0,1] \subset \mathcal{R}$. Let $\mathcal{G}^\star$ be a collection of subsets of $\Gamma^{j\star}$, then the tuple $(\Gamma^{j\star},\mathcal{G}^\star)$ forms a $\sigma$-algebra. The interval, $[0,1]$ is Borel-measurable and the probability measure, $\lambda(\cdot)$ in this case is the Lebesgue measure; the triple, $(\Gamma^{j\star},\mathcal{G}^\star,\lambda)$ forms a probability space with $\lambda(\Gamma^{j\star})=1$, and $\lambda(\phi)=0$. From our previous argument based on the countability and uncountabilty of `right' and `wrong' trajectories, we can infer that the `right' trajectories in $\Gamma^{j}$ will cluster around $\{0\} \in [0,1]$, and would be countable in number, whereas the `wrong' trajectories would cluster around $\{1\} \in [0,1]$, and would be uncountably many. The subset, $R$ in the transformed sample space, $\Gamma^{j\star}$ is denoted by $R^{\star}$. Similarly, the subset $W$ is denoted by $W^{\star}$, and $\{0\}\in R^\star$ and $\{1\}\in W^\star$. On assigning the probability measure to the sets, $R^\star$ and $W^\star$, we observe $\lambda(R^{\star})=0$, since $R^{\star}$ is countable, and $\lambda(W^{\star})=1$, as $W^{\star}$ is uncountable \cite{billingsley2008probability}. What we just proved above is that no matter what, the probability that a `wrong' trajectory is chosen is one, whereas the probability of choosing a `right' trajectory is zero, which in fact, supports Murphy's statement!
\begin{figure}[b]
\begin{center}
\includegraphics[width=0.7\textwidth]{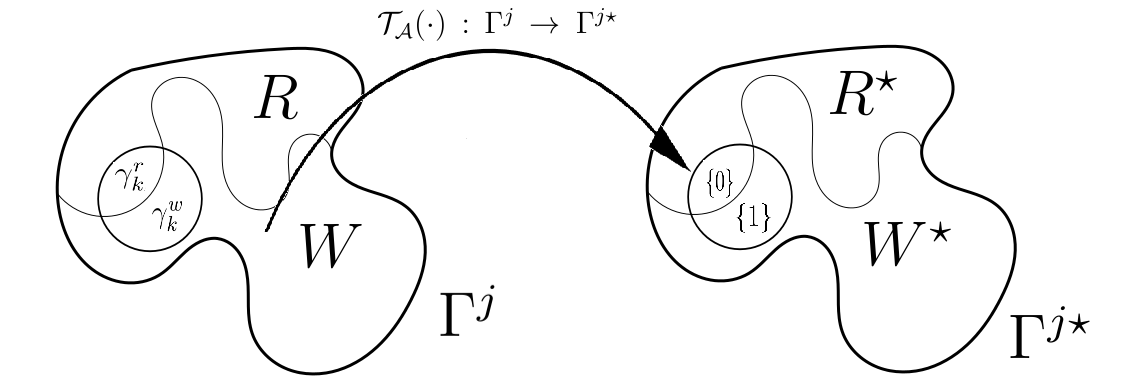}
\end{center}
\caption{Figure shows a homeomorphic relation between the two sets, $\Gamma^j$ and $\Gamma^{j\star}$.}
\end{figure}
There is a fallacy in the above argument. When we say there is a finite probability of any event to happen in some way, there is \emph{also} a finite probability that the event might happen in some other way, that we and also Murphy's statement ignore. As with any natural process, it can happen in infinitely many ways (as seen above), and the \emph{few} so-called `correct' trajectories get `lost' in the multitude of the uncountably many `wrong' ones, in spite of having finite chances of occurring.
\subsection{Dis-proving Murphy's Statement}
\noindent We define two sequences, $\{\gamma^{r}_{n}\} \in R$ and $\{\gamma^{w}_{n}\} \in W$. The sequence, $\{\gamma^{r}_{n}\}$ lies in the neighborhood of $\{0\} \in R^\star$, and the sequence, $\{\gamma^{w}_{n}\}$, in the neighborhood of $\{1\} \in W^\star$ ($\mathcal{T}_\mathcal{A}(\{\gamma^r_n\}\rightarrow\gamma^r) = 0$ and $\mathcal{T}_\mathcal{A}(\{\gamma^w_n\}\rightarrow\gamma^w)= 1$). From our initial assumption, $\mathcal{P}(\gamma^{r}_{k})$ and $\mathcal{P}(\gamma^{w}_{k})$ are both finite for \emph{every} $k$ in the respective sequences, $\{\gamma^{r}_n\}$ and $\{\gamma^{w}_n\}$. Based on our definition for `right' and `wrong' events, the terms in the sequences, $\gamma^{w}_{k}$ and $\gamma^{r}_{k}$ will again be infinitely many. Following our intuitive idea, we can easily foresee the probability, for every element in the sequence, $\gamma^{w}_{k}$ takes the form, $\mathcal{P}(\{\gamma^{w}_{n}\}) \sim n^{-\alpha}, \alpha \geq 1$, for every $\gamma^{w}_{k} \in \{\gamma^{w}_{n}\}$ ($n^{-\alpha}$ is not a distribution function, strictly because for some processes, the probability might decay, as $n^{-2}$, while for others, as $n^{-4}$, and so on). We can see that however small the probability for any process to occur along a specific trajectory, $\gamma^{w}_{k}$ be, it is finite. Since the probability for every trajectory is decaying with an increment in $k$, the summation of the probabilities, $\sum^{\infty}{\mathcal{P}(\gamma^{w}_{k}) < \infty}$ (because as we go nearer to the trajectories in the neighborhood of $\{1\}\in W^\star$, their probability will decrease at an even faster rate, making the probability of the sum of infinitely many such trajectories finite). From the Borel-Cantelli lemma, we can say that the probability of occurrence of these trajectories infinitely many times will be zero, and there are infinitely many of them \cite{billingsley2008probability}. So, the probability of their occurrence is zero. Conversely, for the sequence, $\{\gamma^{r}_{n}\}$, the probability distribution takes the form of $\mathcal{P}(\{\gamma^{r}_{n}\}) \sim n^{-\beta}$, where $\beta \sim 1$ for some $n$, while the probability is \emph{exactly} $1$ for finitely many trajectories. Thus, the summation of probabilities in this case, $\sum^{\infty}{\mathcal{P}(\gamma^{r}_{k})}$ diverges because $n^{-1}$ diverges. From the converse of the lemma, the probability of occurrence of these trajectories infinitely often will be one, and there are infinitely many of them and that such a trajectory (from the set, $R$) will be chosen \emph{almost surely}. We reflect on our findings based on the logic behind Murphy's statement, $\frac{p \rightarrow q, \enspace \neg q}{\therefore, \enspace \neg p}$ and observe that although `wrong' trajectories have finite probability, they do not occur, which is a contradiction to the proposition that, $\neg q \Rightarrow \neg p$ (with finite probabilities). 

\section{Discussion and Conclusion}
\noindent Murphy's statement often serves as an anecdote in many real-life circumstances. Due to its affinity towards the negative aspects of any situation, it has often been misinterpreted to hold a resemblance with the Second Law of Thermodynamics. In this paper, we actually tried to relate both these statements and found quite striking results. As we mentioned earlier, Murphy's statement holds several philosophical connotation,s and appears as a generalized take on various things happening around us all the time. In order to test its scientific validity, we need to construct it as a chain of mathematically consistent statements and derive some logical inference from them. Any proof is invalid if the statements to be proved or disproved are themselves logically incoherent and mathematically not sound. Once this is achieved (in section 2.1), we need to outline a proper proof strategy. This is where the complexity associated with Murphy's statement increases, the reasons being: (\emph{i}) it deals with physical, observable processes, (\emph{ii}) the outcomes of the processes could be anything, and depending upon external perturbations the outcomes and even the processes may change their course of action, and (\emph{iii}) every outcome is plausible, \emph{i.e.}, every process has a finite chance of occurring and along numerous pathways leading to numerous finite possible outcomes. In order to deal with these uncertainties, we made assumptions like: any process or any outcome always has some energy associated with it, and any observable outcome can be achieved in several ways, which we call trajectories or pathways. In order to deal with a complex situation like this, we look for inspiration in the most fundamental laws of nature, and thereby propose that those outcomes, which are thermodynamically feasible shall be achieved along those trajectories that minimize action and maximize entropy. Using mathematical tools from probability theory and logic, we claim to disprove Murphy's statement in this paper. During the course of disproving the statement, we clearly showed the fallacy which is often the mis-interpreted truth behind the statement (section 3.1). We saw that there always exists an equivalent finite chance of an event to occur in different ways when it occurs in a certain way, which Murphy's statement generally ignores. We would further like to add that the intensifier, `definitely' in the construction of Murphy's statement enforces the fact that if something can go wrong, it will go wrong for sure. But we have seen earlier that the probability of something happening along the wrong way, as well as along the right way is finite, which raises a situation of \emph{post-hoc} fallacy in the statement. This type of logical fallacy is known as \emph{post-hoc ergo propter hoc} \cite{damer2012attacking}. A similar situation is observed in the case of the Law of Large Numbers and the Central Limit Theorem. According to the Law of Large Numbers (LLN), given sufficiently large number of trials, the average of the results thus obtained, shall converge (almost surely) to the expected value or mean of the experiment. Similarly, the Central Limit Theorem (CLT) states that the arithmetic mean of a sufficiently large number of iterates of independent random variables (each with a well-defined mean and variance) will be approximately normally distributed, regardless of the underlying distribution \cite{billingsley2008probability}. Both CLT and LLN seem to miss out on a crucial piece of information, \emph{i.e}, the information of the \emph{individual} outcomes or the \emph{individual} distributions respectively! Thus, Murphy's statement seems to be a far-fetched generalization of CLT and LLN in daily life. However, we must not forget that Murphy's statement has a physical essence associated with it (\emph{process}), and therefore needs to be analyzed on a different scale. Since it also deals with large number of outcomes and events, a statistical approach does sound good but a statistical formulation often fails to capture certain fine-tuned characteristics of the system. Our methodology, thus holds strong promises in the study of complex systems, where every motion of a particle counts and little perturbations may drive the system from an organized state to a state of chaos \cite{bak1997nature, bak1987self, heylighen2001cybernetics}. During the formulation of Murphy's statement, we also focused more on the process and the trajectories, and less on the system and its constituting elements. It is not necessary for the system or a system element to be a physical particle, rather it is immaterial and irrespective of the physical definition (or topology) of the system for our reasoning to hold. By a process, we refer to the change in state of the system, which is the same as the change in outcome(s) due to the system undergoing a process in time. As far as our assumption -  any process is associated with energy flows on the energy landscape - holds good, the central idea of the paper will hold true. Further, we mathematically prove that nature orders the occurrence of processes and their outcomes based on energy and entropy considerations. Thus a reinterpretation of Murphy's statement would read as: irrespective of all the possible outcomes and their qualitative aspects of `rightness' and `wrongness' anything that shall happen will definitely happen along the way that must minimize action.  

\section{Acknowledgment}
\noindent The author would like to thank the reviewers for their valuable feedback highlighting the merits and drawbacks of this paper. Their constructive criticism and suggestions have made this paper much more informative and interesting to read. The author would also like to thank his colleagues, Sheetal Surana and Catherine Nipps, for their inputs and lengthy discussions on the topic of this paper.

\end{document}